\documentclass[conference, 10pt]{IEEEtran}
\IEEEoverridecommandlockouts
\usepackage{cite}
\usepackage{amsmath,amssymb,amsfonts}
\usepackage{algorithm, algorithmic}
\usepackage{graphicx}
\usepackage{listings}
\usepackage{textcomp}
\usepackage{multicol, multirow}
\usepackage{xcolor}
\usepackage{url}
\usepackage{caption}
\usepackage{subcaption}
\usepackage[final]{changes}
\usepackage{tcolorbox}
\usepackage{CJKutf8}

\def\BibTeX{{\rm B\kern-.05em{\sc i\kern-.025em b}\kern-.08em
    T\kern-.1667em\lower.7ex\hbox{E}\kern-.125emX}}
\begin{document}
\begin{CJK*}{UTF8}{gbsn}

\newcommand{\fname}{WideSA}
\title{WideSA: A High Array Utilization Mapping Scheme for Uniform Recurrences on ACAP
}
\author{
Tuo Dai,
Bizhao Shi, 
Guojie Luo \\
\{daitoto, bshi, gluo\}@pku.edu.cn \\ 
\textit{School of Computer Science, Peking University}\\\textit{Center for Energy-efficient Computing and Applications, Peking University} 
}

\maketitle

\begin{abstract}
The Versal Adaptive Compute Acceleration Platform (ACAP) is a new architecture that combines AI Engines (AIEs) with reconfigurable fabric.
This architecture offers significant acceleration potential for uniform recurrences in various domains, such as deep learning, high-performance computation, and signal processing. 
However, efficiently mapping these computations onto the Versal ACAP architecture while achieving high utilization of AIEs poses a challenge.

To address this issue, we propose a mapping scheme called \fname, which aims to accelerate uniform recurrences on the Versal ACAP architecture by leveraging the features of both the hardware and the computations.
Considering the array architecture of AIEs, our approach utilizes space-time transformations based on the polyhedral model to generate legally optimized systolic array mappings. 
Concurrently, we have developed a routing-aware PLIO assignment algorithm tailored for communication on the AIE array, and the algorithm aims at successful compilation while maximizing array utilization.
Furthermore, we introduce an automatic mapping framework. This framework is designed to generate the corresponding executable code for uniform recurrences, which encompasses the AIE kernel program, programmable logic bitstreams, and the host program.
The experimental results validate the effectiveness of our mapping scheme. 
Specifically, when applying our scheme to matrix multiplication computations on the VCK5000 board, we achieve a throughput of 4.15TOPS on float data type, which is 1.11$\times$ higher compared to the state-of-the-art accelerator on the Versal ACAP architecture.

\end{abstract}


\section{Introduction}
Modern heterogeneous FPGA architectures, like AMD/Xilinx Versal Adaptive Compute Acceleration Platform (ACAP)~\cite{xilinx:acap}, combine AI Engines (AIEs) with programmable logic (PL) to boost applications in the AI and intelligent signal processing domains. 
In these domains, uniform recurrences~\cite{karp67:ure}, which comprise nested loops with uniform dependencies, are prevalent types of computations. 
Regrettably, there is currently a lack of established development methodologies for efficiently mapping large-scale uniform recurrences onto the Versal ACAP architecture with high utilization of AI Engines.

The ACAP architecture comprise an array of several hundred AIE cores, such as $8\times50$ in the VC1902 architecture~\cite{ahmad2019:vc1902}, interconnected through a mesh network-on-chip (NoC). 
Each AIE core consists of vector processing and load/store units, functioning as a very-long-instruction-word (VLIW)~\cite{fish83:vliw} processor to deliver high-performance vectorized computations. 
To facilitate communication among the AIE cores, the NoC is utilized for inter-core communication, enabling efficient data transfers between cores. 
Moreover, neighboring cores utilize shared buffers, providing higher bandwidth for data exchange. 
When it comes to data transfer to and from the AIEs, there are hundreds of I/O ports available, supporting terabytes of bandwidth.

As ACAP demonstrates a remarkable capacity for intense computation, developing acceleration designs on the architecture has become an urgent trend in recent times. 
However, current efforts have not succeeded in achieving high utilization of the AIE array.
For example, Vitis-AI~\cite{xilinx:vitis-ai} introduces the DPU~\cite{jia22:dpu} for the VC1902 architecture, but only accomplishes a 64\% AIE utilization. 
There are several ongoing challenges associated with developing designs with high array utilization on the Versal ACAP architecture:
\begin{itemize}
    \item \textbf{Increased programming complexity:} Higher AIE utilization results in more cores that need to be programmed with certain intrinsics. In some situations, different cores execute different programs, necessitating significant human effort to develop such accelerators.
    \item \textbf{Increased placement and routing difficulty:} Mapping computations onto the Versal ACAP architecture with high utilization of AIEs often necessitates careful placement and routing of AIEs and data communications. 
    From the perspective of AIE compilation, attaining high AIE utilization typically results in difficulties in placing cores and buffers, as well as routing streaming communications on the NoC.
    For example, CHARM~\cite{zhuang2023:charm} struggles to compile large designs on Vitis 2022.1.
    \item \textbf{Extended compilation time:} The default compilation tools provided by AMD/Xilinx Vitis employ ILP algorithms to find placement and routing solutions. Consequently, a larger number of cores results in a longer time to find a legal solution.
\end{itemize}

To address these challenges, we propose \fname, a high array utilization mapping scheme for uniform recurrences on the Versal ACAP architecture.
By leveraging the AIE array architecture, we apply space-time transformation and loop nest transformation using the polyhedral model, generating systolic-like mappings on the AIE array. 
On one hand, systolic designs assign similar workloads to different cores, enabling us to reuse a single core program and thereby reduce human effort.
On the other hand, systolic designs regularize both the placement and communication of cores, simplifying the placement and routing process.
Additionally, we designed a routing-aware PLIO assignment algorithm to improve the success rate of compilation. 
We also developed an automatic framework to generate the corresponding code for heterogeneous backends, including AIEs, PL, and host.
In the evaluation section, we demonstrate the effectiveness of \fname ~by successfully implementing executable acceleration systems for various uniform recurrences, accommodating different data types. 
Our approach achieves high throughput with high utilization of AIEs.

We summarize our contributions as follows:
\begin{itemize}
    \item We propose a mapping scheme, based on the polyhedral model, for uniform recurrences that generates a systolic design on ACAP with high AIE utilization.
    \item We design a routing-aware PLIO assignment algorithm that takes into account the characteristics of systolic mappings and the AIE architecture, thereby facilitating an efficient compilation process.
    \item We develop an automatic framework that generates corresponding code for heterogeneous backends based on the mapping results.
    \item We achieve high throughput across different computations and data types, outperforming state-of-the-art methods.
\end{itemize}

\section{Background}
\label{sec:back}
\subsection{Versal ACAP Architecture and Workflow}
\subsubsection{Hardware Features}
\begin{figure}
    \centering
    \includegraphics[width=0.45\textwidth]{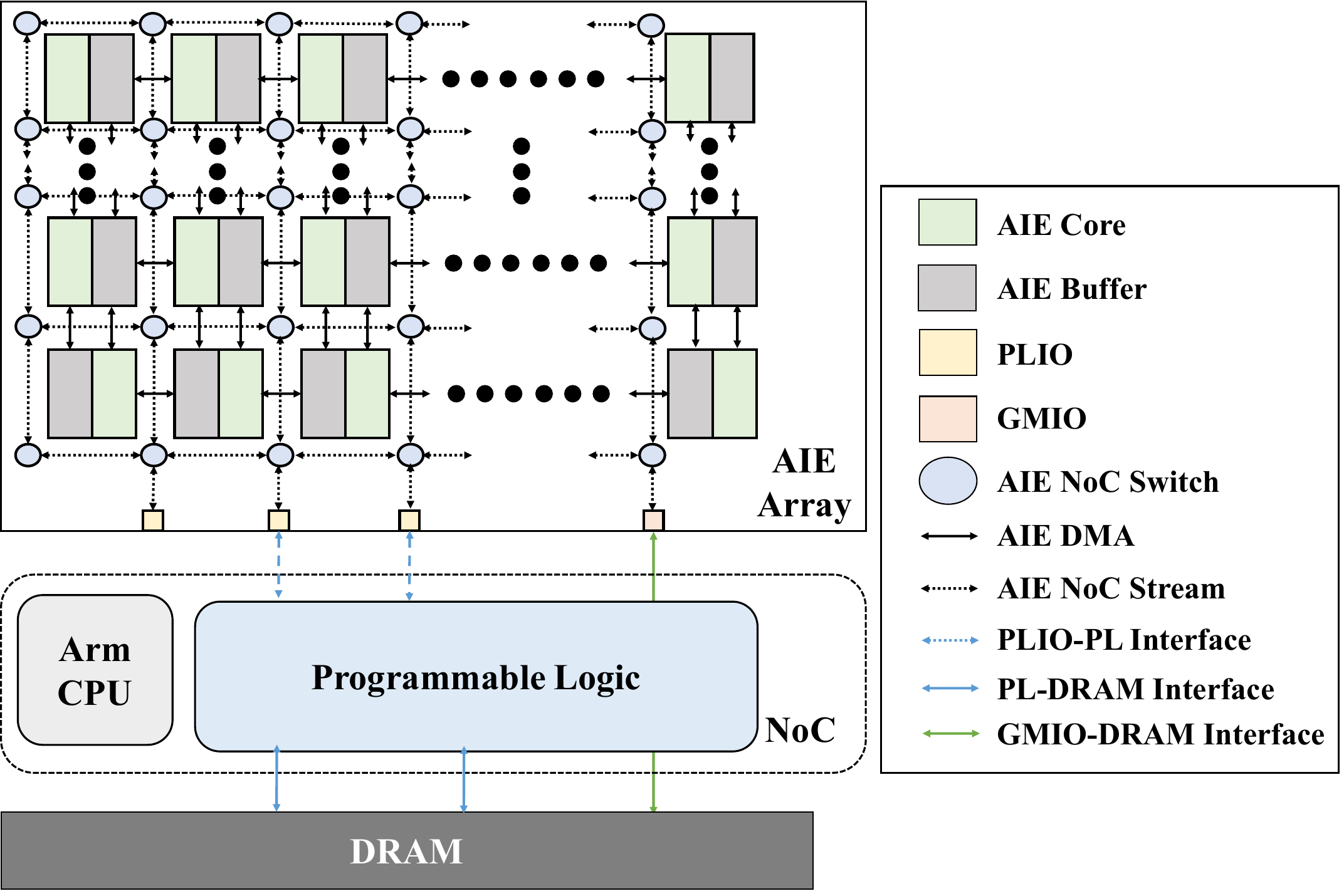}
    \caption{Versal ACAP Architecture}
    \label{fig:acap-arch}
\end{figure}


AMD/Xilinx has developed the Versal ACAP architecture to cater to the increasing demands of next-generation wireless processing and machine learning applications. 
Figure~\ref{fig:acap-arch} illustrates the detailed architecture of VCK5000, an evaluation kit for Versal ACAP, comprising the CPU, PL, and AIE components.
The AIE array on VCK5000 consists of $8 \times 50$ AIE cores, with each core capable of generating 128 MACs of int8 data type every cycle at a frequency of 1 GHz or higher.
Moreover, the AIE cores operate in single-instruction-multiple-data (SIMD) mode using a VLIW pattern, enabling acceleration of a large number of vectorized computations.

In Figure~\ref{fig:acap-arch}, we identify five data transfer methods, including those within the AIE array and among the AIE, PL, and DRAM components.
These methods are referred to as \textbf{AIE DMA}, \textbf{AIE NoC stream}, \textbf{PLIO-PL}, \textbf{PL-DRAM}, and \textbf{GMIO-DRAM} interfaces.
We profile these data transfer methods on VCK5000 and present the results in Table~\ref{tab:band-prof}.
Within the AIE array, each AIE core has direct memory access (DMA) ports connected to four neighboring local buffers with a width of 256 bits.
Using the \textbf{AIE DMA} method, a total data transfer rate of up to 15.6 TB/s can be achieved.
Furthermore, each AIE core is linked to the NoC through a stream interface with a width of 32 bits.
The data transfer bandwidth through the \textbf{AIE NoC stream} method reaches a maximum of 2 TB/s, which is lower compared to the DMA method.
The PLIO ports, responsible for data communication between the PL and AIE array, can provide a maximum bandwidth of 1.52 TB/s.
Based on the profiling results, utilizing the \textbf{AIE DMA} method for data transfer proves beneficial in overcoming communication bottlenecks, aligning with the dataflow in systolic array designs.
In terms of data communication with DRAM, the bandwidth is approximately 0.1 TB/s, significantly lower than the on-chip data transfer methods.
This observation inspires us to exploit data locality within computations to enhance overall performance.

\begin{table}[]
    \setlength\tabcolsep{4.0pt}
    \centering
    \caption{Different Data Communication Bandwidth on the Versal ACAP Architecture}
    \label{tab:band-prof}
    \begin{tabular}{l|cccc}
    \hline
     \multicolumn{1}{c|}{\textbf{Methods}} & \textbf{Frequency} & \textbf{Bitwidth} & \textbf{Channels} & \textbf{Total} \\ \hline
    \textbf{AIE DMA} &  1.25 GHz & 256 bits & 400 & 15.6 TB/s   \\
    \textbf{AIE NoC Stream} &  1.25 GHz &  32 bits & 400 & 1.95 TB/s   \\
    \textbf{PLIO-PL} &  1.25 GHz &  128 bits & 78 & 1.52 TB/s    \\
    \textbf{GMIO-DRAM} &  1.25 GHz &  64 bits & 16 & 0.125 TB/s   \\
    \textbf{PL-DRAM} &  0.50 GHz  & -  & 4 & 0.100 TB/s \\ \hline
    \end{tabular}
    
\end{table}

\subsubsection{Software Programming Model}
AMD/Xilinx offers a development tool for AIEs and Versal ACAP integrated into Vitis.
The programming model~\cite{xilinx:aie-prog} designed for AIEs consists of two levels: a graph program across the AIE array with each node representing an AIE kernel program.
The graph program represents the dataflow information among AIE kernels and between the AIE and I/O ports.
The compiler in Vitis transforms the dataflow graph into a subnetwork of physical AIE cores, determines the placement of buffers, and configures NoC stream routing.
Since placement and routing are NP-hard problems, the compiler employs ILP solvers to process these two phases.
However, as the design scale increases and AIE utilization becomes high, finding a legal solution efficiently becomes challenging for the solvers~\cite{cook11:ilp}.
To address this, incorporating constraints for placement and routing helps alleviate the congestions and accelerates the solvers in finding solutions.
The systolic design scheme provides a regular pattern for placement and routing, which is suitable for constructing these constraints.


\subsection{Uniform Recurrences and Systolic Array Mapping}
Uniform recurrences refer to computations that consist of nested loops, where all dependencies are uniform. 
These types of computations are commonly found in AI and signal processing applications, such as matrix multiplication, 2D convolution, FIR filtering, and so on. 
Several prior works~\cite{cong18:polysa, wang2021:autosa, lai2020:susy} have focused on generating systolic array designs for uniform recurrences on FPGAs, employing the polyhedral model for loop transformations to explore successful mappings. 
The polyhedral model~\cite{ben10:poly, bond13:affine} serves as a compilation framework for loop transformation, encompassing space-time transformation, latency hiding, SIMD vectorization, fusion, and more. 
A legal combination of these transformations represents a schedule within the polyhedral model, and the goal of systolic design mapping is to find the optimal schedule.

An AIE kernel handles more computations compared to a PE in typical systolic arrays. Additionally, specific hardware features of the AIE array differ from those of common systolic arrays. As a result, the mapping problem on the Versal ACAP architecture is not a straightforward systolic array mapping.
Consequently, 
it is necessary to model corresponding transformations and constraints within the polyhedral model, an area that has not yet been extensively researched.

\section{Systolic Mapping Scheme on ACAP}
\label{sec:method}
\subsection{Kernel Scope Demarcation}
\begin{figure}[b]
    \centering
    \includegraphics[width=0.48\textwidth]{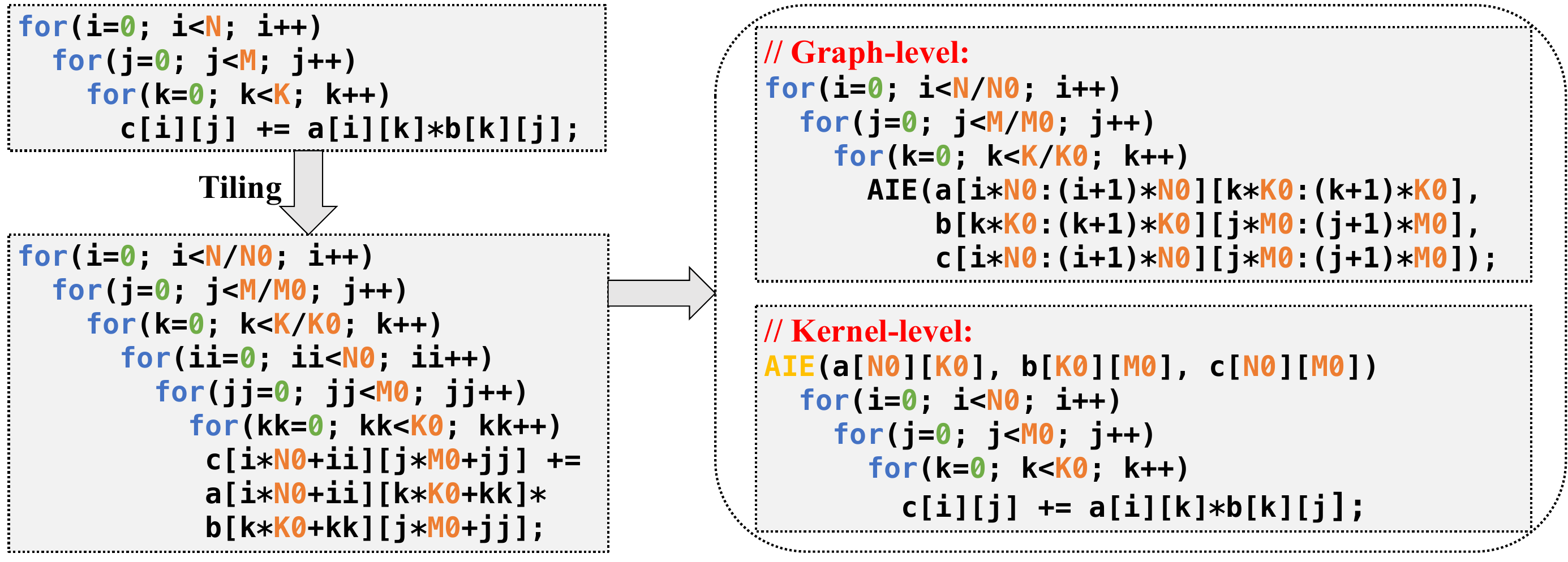}
    \caption{Kernel Scope Demarcation}
    \label{fig:two-level}
\end{figure}
\begin{figure}[t]
    \centering
    \includegraphics[width=0.48\textwidth]{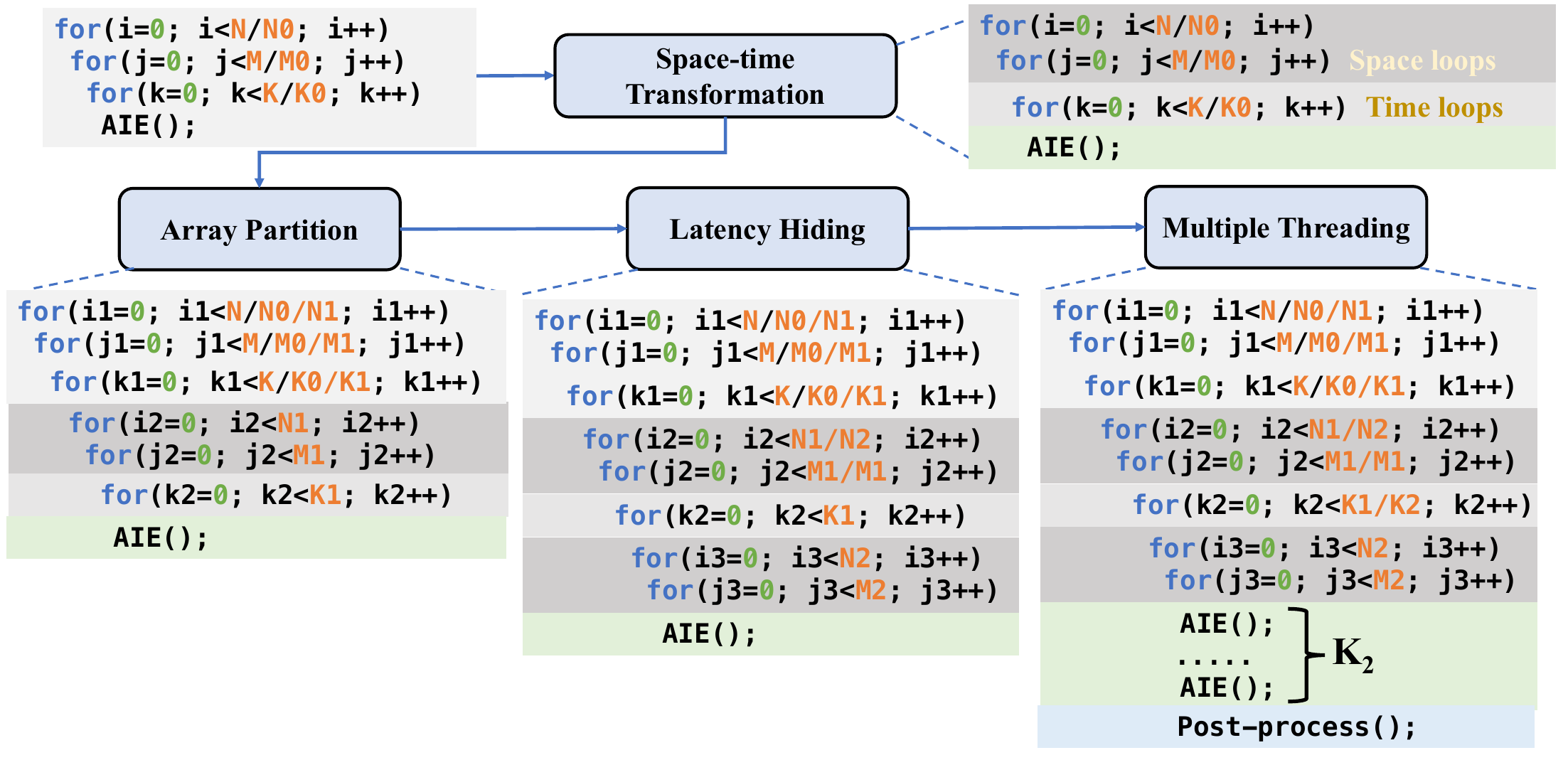}
    \caption{Polyhedral Model-Based Systolic Mapping}
    \label{fig:trans}
\end{figure}
According to the programming model of AIEs, it is necessary to demarcate the scope of codes mapped to execute on a single AIE core and the outer loop nests to be mapped to the AIE array. 
This demarcation allows us to decompose the mapping problem into graph-level mapping and kernel-level mapping, which are independent of each other after selecting tiling factors. 

Polygonal tiling~\cite{and2003:tiling, ross23:tiling}, an effective solution for workload partitioning in uniform polyhedral domains, plays a crucial role in determining the innermost and outer loop nests for tiling. 
We illustrate the tiling process using the MM example with ($N_0, M_0, K_0$) as the tiling factors, involving loop re-indexing, tiling, and rewriting, as depicted in Figure~\ref{fig:two-level}.
Building on prior works, we consider the specific features of the AIE array when performing the demarcation.



\subsection{Systolic Mapping Generation}
To generate systolic array designs on the AIE array following kernel scope demarcation, we utilize the polyhedral model, drawing inspiration from AutoSA~\cite{wang2021:autosa}, to facilitate loop transformations. 
To be specific, we employ four types of transformation techniques, as depicted in Figure~\ref{fig:trans}.

\subsubsection{Space-time Transformation}
The first step involves performing space-time transformation to map the graph-level loop nests to a systolic array design. 
We identify loops in the outermost loop band with dependence distances no greater than one and consider them as candidate space loops. 
Subsequently, we enumerate all possible combinations of space loops from the candidate pool. 
The selected space loops are then permuted in the outermost position, while the loops below them are designated as time loops.
Due to the constraints imposed by the hardware shape of the AIE array, the mapper generates only 1D and 2D systolic arrays. 
This step results in the generation of multiple systolic arrays, each with a unique schedule. 
As shown in Figure~\ref{fig:trans}, we choose loops $i$ and $j$ as the space loops (on dark gray background) and loop $k$ as the time loop (on light gray background) in the MM example.

\subsubsection{Array Partition}
To accommodate the limited number of AIEs in the horizontal and vertical directions of the AIE array, array partitioning becomes necessary when mapping a large array. 
In order to achieve this, we apply tiling to the outermost permutable loop that contains the space loops. 
In Figure~\ref{fig:trans}, we illustrate an example where we tile the outermost loop band in the MM example using the tiling factors $(N_1, M_1, K_1)$. 
The point loops originating from the original loops are retained as the space loops. 
This results in a 2D systolic array with dimensions of $N_1 \times M_1$ (on dark gray background).

\subsubsection{Latency Hiding}
Latency hiding plays a crucial role in mitigating the pipeline stalls caused by loop-carried dependencies in computational statements. 
In the case of the MM example, the accumulate operations in the statement introduce loop-carried dependence within the loop, resulting in long latency in the systolic chain. 
To address this issue, we identify parallel loops in the polyhedral model schedules, applies tiling to these loops, and permutes the point loops to the innermost position. 
As an illustration, loops $i$ and $j$ are identified as parallel loops in the MM example. 
We extract them using the tiling factors ($N_{2}$, $M_{2}$) and permute the point loops to the innermost position. 
Since there are no loop-carried dependencies on the innermost loop, the latency of design reduce as the chain length shortened.

\subsubsection{Multiple Threading}
As AIE cores execute concurrently, the AIE array inherently supports multiple threading. 
Leveraging this characteristic, utilizing multiple AIEs to execute the same instructions but different indexing can significantly enhance overall performance. 
We identify parallelizable loops in the time loops that do not have data dependence.
In the MM example, the loop $k$ is identified as a parallelizable loop. 
We can apply tiling to this loop using the factors $K_2$.
The point loop is permuted to the innermost position and completely unrolled to generate multiple threads of AIEs. 

\subsection{Placement and Routing Constraints Construction}
The systolic design generated in the previous section represents an abstract mapping scheme. 
Consequently, it is essential to use the space loops as input and generate an actual mapped graph for AIE array that considers placement and routing constraints. 
The mapped graph consists of nodes, representing AIE cores and input/output ports, and edges, which connect the ports of the nodes. 
The placement and routing constraints involve assigning coordinates to the AIE cores, buffers, and input/output ports, as well as determining the routing paths for the edges.
In the subsequent subsections, we introduce the graph builder and routing-aware PLIO assignment, which are responsible for constructing the mapped graph and generating the associated placement and routing constraints, respectively.
\subsubsection{Graph Builder}
\begin{figure}
    \centering
    \includegraphics[width=0.4\textwidth]{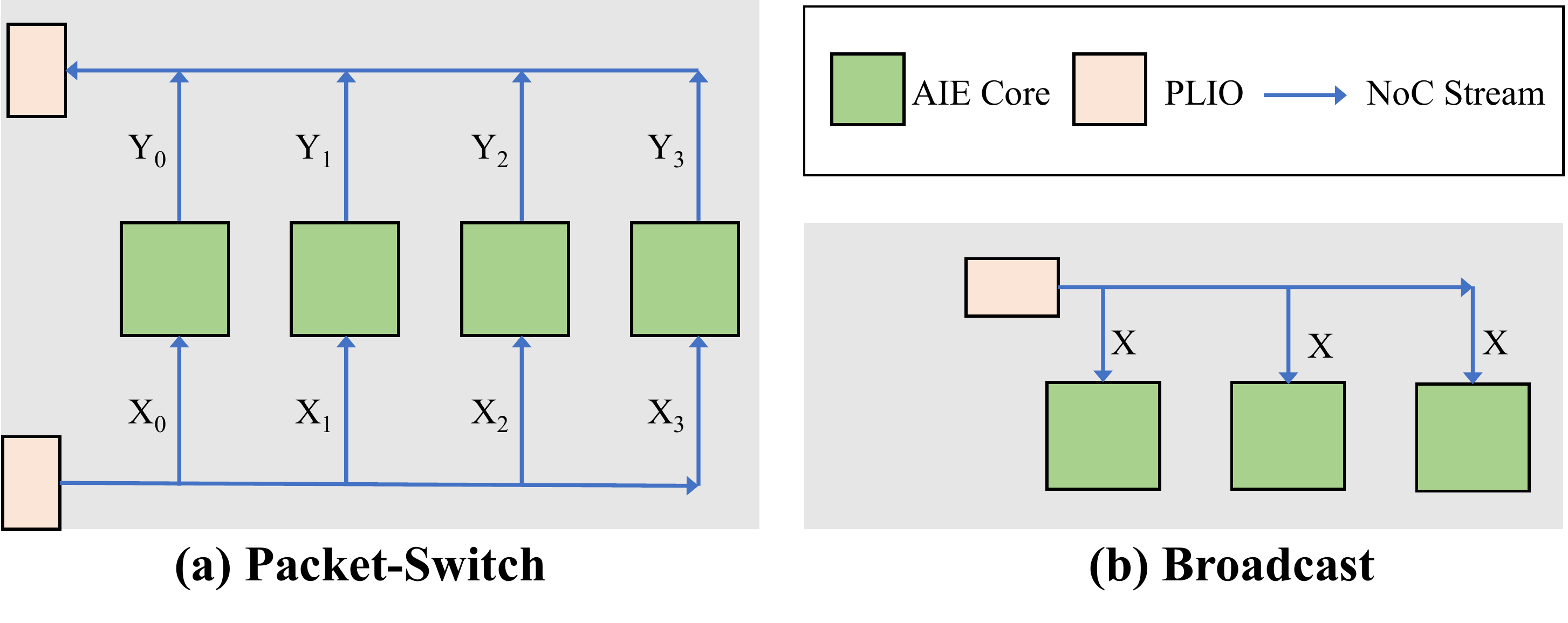}
    \caption{Communication Methods for PLIO Ports Utilization Reduction}
    \label{fig:communi}
\end{figure}

To construct the mapped graph, we iterate through all coordinates in the space loops and create a node for each pair of coordinates in the 2D systolic array, representing an AIE core.
Next, we identify the data communications between AIE cores based on the dependencies within the space loops. 
Following the definitions in AutoSA~\cite{wang2021:autosa}, there are three types of data dependences:
\begin{itemize}
    \item Read dependence: Transfer the read-only data.
    \item Flow dependence: Transfer the intermediate data.
    \item Output dependence: Transfer the output-only data.
\end{itemize}

Based on these data dependences and the space loops, we define the I/O ports and edge directions. 
Since AIEs do not support intermediate results between different iterations, we treat flow dependences as input dependencies when constructing I/O ports.
The polyhedral model for the array access to matrix $A$ in the MM recurrences is $\{i, j, k\} \rightarrow \{i, j{+}1, k\}$, and when loops $j, k$ are the space loops, the direction is $(1, 0)$. 
We connect the input ports from the corresponding nodes with a constant and non-zero distant direction.

As for the output ports, the boundary input ports, and the zero distant direction ports, we create PLIO ports as the other end of the connection edge. 
To adhere to the limitation on the number of PLIO ports, we utilize packet-switch communications and broadcast communications to reduce the number of used ports, as depicted in Figure~\ref{fig:communi}.

\subsubsection{Routing-Aware PLIO Assignment}
Once we have the mapped graph, we search for available cores on the AIE array to place the AIE kernels. 
To facilitate efficient communication between neighboring cores, we assign the buffers of ports connecting these cores to the shared buffer of the cores, forming part of the placement constraints. 
These constraints enable the transformation of the kernels' placement into a regular duplicate pattern of a single kernel.

Aside from facilitating neighboring communication, it is necessary to construct paths between PLIO ports and AIE cores for data input and output.
Considering the mesh structure of the NoC on the AIE array, and given that PLIOs are always located in Row 0, we can compute the routing congestion by counting the horizontal data transfer numbers. 
For instance, we compute the congestion for the \textit{west} direction as follows:
\begin{equation*}
    \textit{Cong}^{\textit{west}}_{i} = \sum_{p\in \text{PLIOs}, x \in \text{AIEs}} W_{i}[p][x],
\end{equation*}
\begin{equation*}
    W_{i}[p][x] = \left\{
    \begin{array}{cl}
        1 & \text{($p_\text{col}{<}i$ and $x_{\text{col}}{>}i$ and $(x,p)\in\text{Edges})$ or} \\
          & \text{($p_\text{col}{>}i$ and $x_{\text{col}}{<}i$ and $(p,x)\in\text{Edges}$)} \\
        0 & \text{Otherwise}
    \end{array}
    \right.
\end{equation*}
where $p_\text{col}$ and $x_\text{col}$ represent the column coordinates of PLIO $p$ and AIE $x$, respectively.

The computation of the congestion for the \textit{east} direction is symmetrical.

Consequently, the routing challenges essentially transform into issues of PLIO assignment. 
We formulate the assignment of PLIO ports as a satisfiability problem subject to routing resource constraints.
We check if there exists a set of values for PLIOs that satisfies the following constraints:
\begin{equation*}
\forall i \in \text{Columns}, \textit{Cong}^{\textit{west}}_{i} \leq \text{RC}_{\textit{west}}, \ \ 
\textit{Cong}^{\textit{east}}_{i} \leq \text{RC}_{\textit{east}}
\end{equation*}
where $\text{RC}_\textit{west}$ and $\text{RC}_{\textit{east}}$ denote the available routing resources in the AIE array.

To seek the feasible assignment of PLIO ports, we employ a heuristic greedy algorithm outlined in Algorithm~\ref{alg:init}.
In this algorithm, we initialize the placement of the PLIO ports by calculating the median value of the row numbers of the connected AIE cores. 
If the initially computed placement coordinate is not available, we search for the nearest available coordinate instead.
This heuristic greedy algorithm balances the routing congestion among the PLIO ports.
By considering the connectivity with the AIE cores, it generates an optimal placement for the PLIO ports, ensuring successful routing on the NoC.
The algorithm takes into account the availability of coordinates and selects the most suitable placement to minimize congestion.

\begin{algorithm}
\caption{Routing-Aware PLIO Assignment Algortihm}
\label{alg:init}
\begin{algorithmic}[1]
\REQUIRE Numbers of PLIO ports $N$, AIE cores $X$
\ENSURE Initialized PLIO assignment set $P$
\STATE Initialization available placement sets $A$ as all columns that have PLIO ports.
\FOR{$i \leftarrow 1$ \TO $N$}
    \STATE $S = [], num = 0$
    \FOR{$x \in X$}
        \IF{$(p, x) \in \text{Edges}$}
          \STATE  $S$.append($x_\text{col}$)
          \STATE  $num += 1$
         \ENDIF
    \ENDFOR
    \STATE Sort $S$ to find the median: sort($S$, $S + num$)
    \STATE $P[i] = \text{find\_nearest}(A, S[num/2])$
    \STATE remove($A$, $P[i]$)
\ENDFOR
\RETURN $P$
\end{algorithmic}
\end{algorithm}

By generating these constraints for the placement and routing of AIE kernels, buffers, and PLIO ports, we can significantly simplify the task for the AIE compiler. 
These constraints provide valuable information and guidelines for the compilers to optimize the placement and routing process, ultimately leading to a high utilization of the AIE array.

\section{Automatic Mapping Framework}
\label{sec:frame}

\begin{figure}[ht]
    \vspace{-0.4cm}
    \centering
    \includegraphics[width=0.48\textwidth]{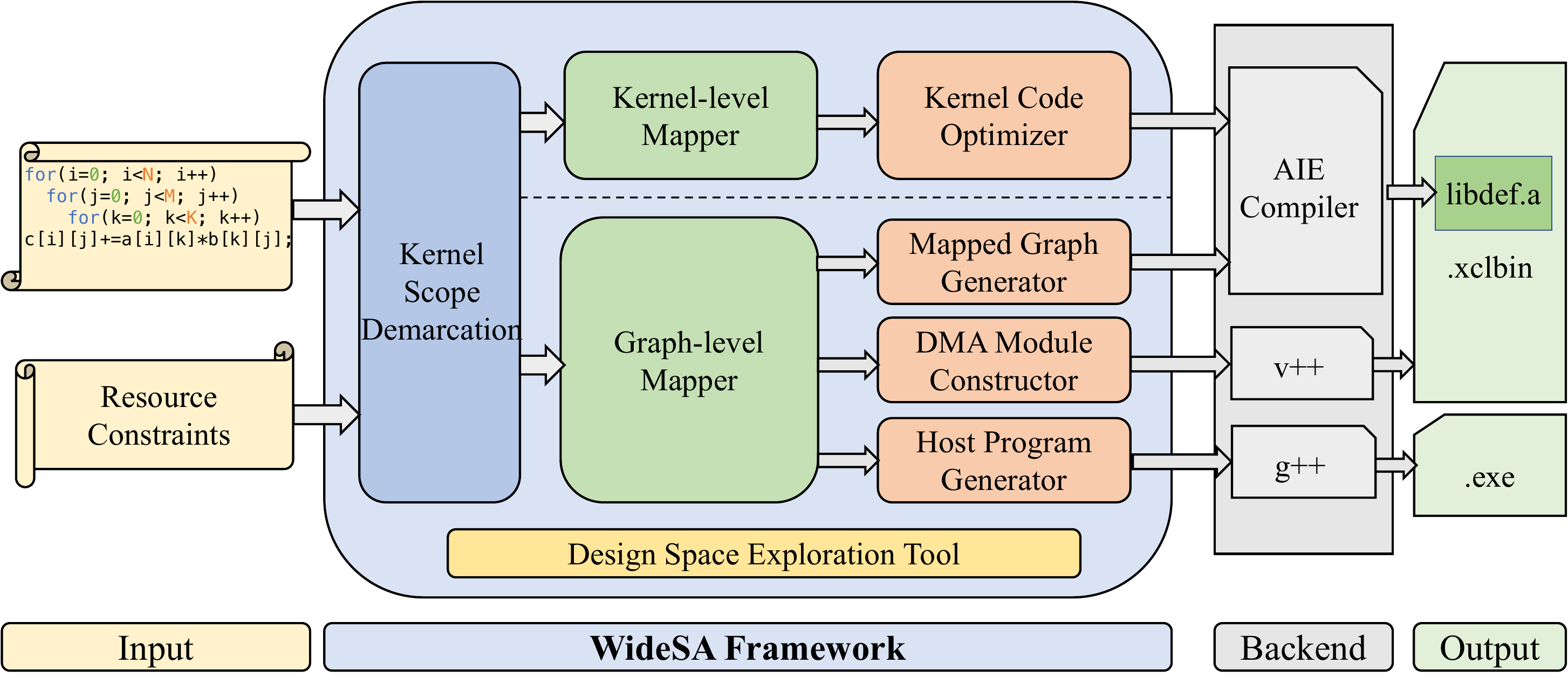}
    \caption{Overview of WideSA Automatic Framework}
    \vspace{-0.4cm}
    \label{fig:wide-frame}
\end{figure}

To facilitate the computation of uniform recurrence, we have developed an automatic mapping framework that implements the full functional modules on the Versal ACAP architecture, as shown in Figure~\ref{fig:wide-frame}.

Specifically, we introduce a kernel-level mapper, a DMA module constructor, and a host program generator, which work in conjunction with the kernel scope and graph mapper described in the previous section.

The kernel-level mapper and optimizer transform the C++ program into a program with AIE intrinsics, leveraging the capabilities of the AIE vector processor to exploit parallelism and optimize performance. 
Moreover, we design the architecture of efficient DMA modules, which serve as the buffers of AIEs on the PL, in the DMA module constructor. 
This architecture is tailored to the characteristics of both the hardware and the computations involved. 
In addition, we engineer a host program generator to generate a controller program that oversees global scheduling.

\section{Evaluation}
\subsection{Benchmark and Experimental Setup}
In this section, we select four representative uniform recurrences with various data types as benchmarks to evaluate the performance of WideSA. 
The selected benchmarks include matrix multiplication (MM), 2D convolution (2D-Conv), 2D Fast Fourier Transformation (2D-FFT), and FIR filter~\cite{Parhi2007:VlsiDS}. 
The problem sizes and corresponding data types are provided in Table~\ref{tab:benchmark}. 
Here, Cfloat refers to the complex float data type and Cint16 refers to the complex 16-bit integer data type.
All the experiments are conducted on VCK5000 with 250 MHz on PL and 1.25 GHz on AIE. 
AMD/Xilinx Vitis 2022.1 is used as the compilation backend tool.

\begin{table}[]
\centering
\caption{Evaluation Benchmarks}
\label{tab:benchmark}
\begin{tabular}{cccc}
\hline
\textbf{Benchmarks} & \textbf{Dimension}& \textbf{Problem Size}  & \textbf{Data Types} \\ \hline
\multirow{4}*{\textbf{MM}} & \multirow{4}*{[i, j, k]} & [8192, 8192, 8192]    &  Float     \\
& &[10240, 10240, 10240] & Int8 \\
& & [9600, 9600, 9600] & Int16 \\
& & [8192, 8192, 8192] & Int32 \\ \hline
\multirow{4}*{\textbf{2D-Conv}} & \multirow{4}*{[h, w, p, q]} & [10240, 10240, 4, 4] & Float \\
& & [10240, 10240, 8, 8] & Int8 \\
& & [10240, 10240, 4, 4] & Int16 \\
& & [10240, 10240, 4, 4] & Int32 \\ \hline
\multirow{2}*{\textbf{2D-FFT}} & \multirow{2}*{[row, col]} & [8192, 8192] & Cfloat \\
& & [8192, 8192] & Cint16 \\ \hline
\multirow{4}*{\textbf{FIR Filter}} & \multirow{4}*{[n, taps]} & [1048576, 15] & Float \\
& & [1048576, 15] & Int8 \\
& & [1048576, 15] & Int16 \\
& & [1048576, 15] & Cfloat \\ \hline
\end{tabular}
\vspace{-0.6cm}
\end{table}

\begin{table*}[]
\caption{Comparison of Throughput and AIE Efficiency on Benchmarks}
\label{tab:perf-tps}
\begin{tabular}{l|l|cccc|cccc|cc|cccc}
\hline
Method & \multicolumn{1}{c|}{Metric} & \multicolumn{4}{c|}{\textbf{MM}} & \multicolumn{4}{c|}{\textbf{2D-Conv}} & \multicolumn{2}{c|}{\textbf{2D-FFT}} & \multicolumn{4}{c}{\textbf{FIR Filter}} \\ \hline
  \multicolumn{2}{c|}{Data type}   &   Float     &   Int8    &   Int16    &   Int32    &  Float      &   Int8    &  Int16     & Int32      &      Cfloat          &     Cint16          &   Float    &  Int8     &   Int16    &   Cfloat    \\ \hline
\multirow{3}{*}{Baseline}      &  \#AIEs     & 384       &   384    &  384     &    384   &  -      &  256     &   -    &   -    &   10             &   10    &   10   &  10     &  10     &  10     \\
                       &   TOPS    & 3.73       &  29.78     &   7.82    &   3.72   &    -    &  31.40 & -               &   -     &   0.04    & 0.13                 &  0.15     &   2.56    &   0.62    &   0.15    \\
                       &  TOPS/\#AIEs     &  0.010      &  0.077     &    0.020 &  0.010   &    -    & \textbf{0.123}      &  -     & -      &   \textbf{0.004}             &  \textbf{0.013}            & \textbf{0.015}      &  \textbf{0.256}     & \textbf{0.062}      & \textbf{0.015}      \\ \hline
\multirow{3}{*}{WideSA}      &   \#AIES    & 400       &   400    &  400     & 400      &  400      & 400      &  400     &   400    &    320            &  320             & 256      &  256     &  256     &   256    \\
                       &  TOPS     &  \textbf{4.15}      &  \textbf{32.49}      &  \textbf{8.10}     &  \textbf{3.92}     &  \textbf{4.50}      &  \textbf{36.02}     &  \textbf{10.35}     & \textbf{4.48}      &   \textbf{1.10}             &   \textbf{3.83}            & \textbf{2.92}      &  \textbf{39.3}     & \textbf{9.47}      & \textbf{2.89}       \\
                       &  TOPS/\#AIEs     & \textbf{0.010}       &  \textbf{0.081}     &  \textbf{0.020}     &   \textbf{0.010}    & \textbf{0.011}       &  0.090     & \textbf{0.025}      & \textbf{0.011}       &         0.003       &   0.012            &  0.012    &  0.100     & 0.037      & 0.011      \\ \hline
\end{tabular}

\vspace{-0.4cm}
\end{table*}

\subsection{Full System Performance}
\label{sec:exp-b}
\begin{table}[]
\setlength\tabcolsep{3.1pt}
    \centering
    \caption{MM Performance Comparison between PL-only and WideSA Design}
    \label{tab:comp-pl}
    \begin{tabular}{l|cccc|cccc}
\hline
 & \multicolumn{4}{c|}{\textbf{PL-only}} & \multicolumn{4}{c}{\textbf{WideSA}} \\ \hline
Data Type & Float    &   Int8  &  Int16   & Int32    & Float    &  Int8   &  Int16   &  Int32  \\
DSPs &  1536   &  1528   &  1516   &  1536  &  152   &   60  &  67   & 65   \\
\#AIEs &  0   & 0    & 0    & 0   & 400   &  400   &  400   &  400  \\
TOPS &  0.59   &  5.77   &  2.16   & 0.60   &  \textbf{4.15}   &  \textbf{32.49}   & \textbf{8.10}    &  \textbf{3.92}  \\
Power (W) & 19.5    & 18.8    & 18.6    &  19.5  &  55.8   & 54.4    & 54.9    & 55.6   \\ 
TOPS/W &  0.03   & 0.31    &  0.12   & 0.03   &  0.07   & 0.60  &  0.15   & 0.07   \\ 
Norm. TOPS/W & 1.00x & 1.00x & 1.00x & 1.00x & \textbf{2.25x} & \textbf{1.94x} & \textbf{1.29x} & \textbf{2.25x} \\ 
\hline
\end{tabular}
\vspace{-0.5cm}
\end{table}

We conducted a comparison of the throughput between WideSA and other state-of-the-art AIE designs for the same problem size.
For the MM benchmark, we successfully compiled the CHARM code~\cite{zhuang2023:charm} for the target VCK5000 with AMD/Xilinx Vitis 2022.1, incorporating placement and routing constraints, as the baseline. 
As for the 2D-Conv benchmark, we selected the released 8-PEs version of Vitis-AI DPU~\cite{xilinx:vitis-ai} which only supports Int8 data type, utilizing 256 AIEs running at 1.33 GHz and the PL at 350 MHz, as the baseline. 
Furthermore, we used the open-source designs from the Vitis DSP Library~\cite{xilinx:dsp-lib} as the baselines for the 2D-FFT and FIR filter benchmarks.

The results presented in Table~\ref{tab:perf-tps} demonstrate that WideSA achieves significantly higher throughput with high utilization of AIEs.
Additionally, we computed the AIE efficiency by considering the throughput and the number of used AIEs. 
The results indicate that WideSA maintains similar efficiency to \cite{zhuang2023:charm} for MM, as both approaches exhibit AIE utilization over 95\%.
When compared to the baselines with lower AIE utilizations, WideSA trades AIE efficiency (TOPS/\#AIEs) for a high overall performance (TOPS) and is bounded by memory bandwidth.

Moreover, we conducted a comparison of the performance and energy efficiency of MM using WideSA and PL-only designs on the VCK5000 target, which has 1968 DSP58 IPs at total. 
For the PL-only designs, we utilize AutoSA~\cite{wang2021:autosa} as the systolic array generator.
The results presented in Table~\ref{tab:comp-pl} demonstrate that our approach achieves up to 2.25$\times$ higher energy efficiency compared to the PL-only designs.

\subsection{Scalability of WideSA on MM examples}
\label{sec:exp-c}
\begin{figure}
    \centering
    \includegraphics[width=0.45\textwidth]{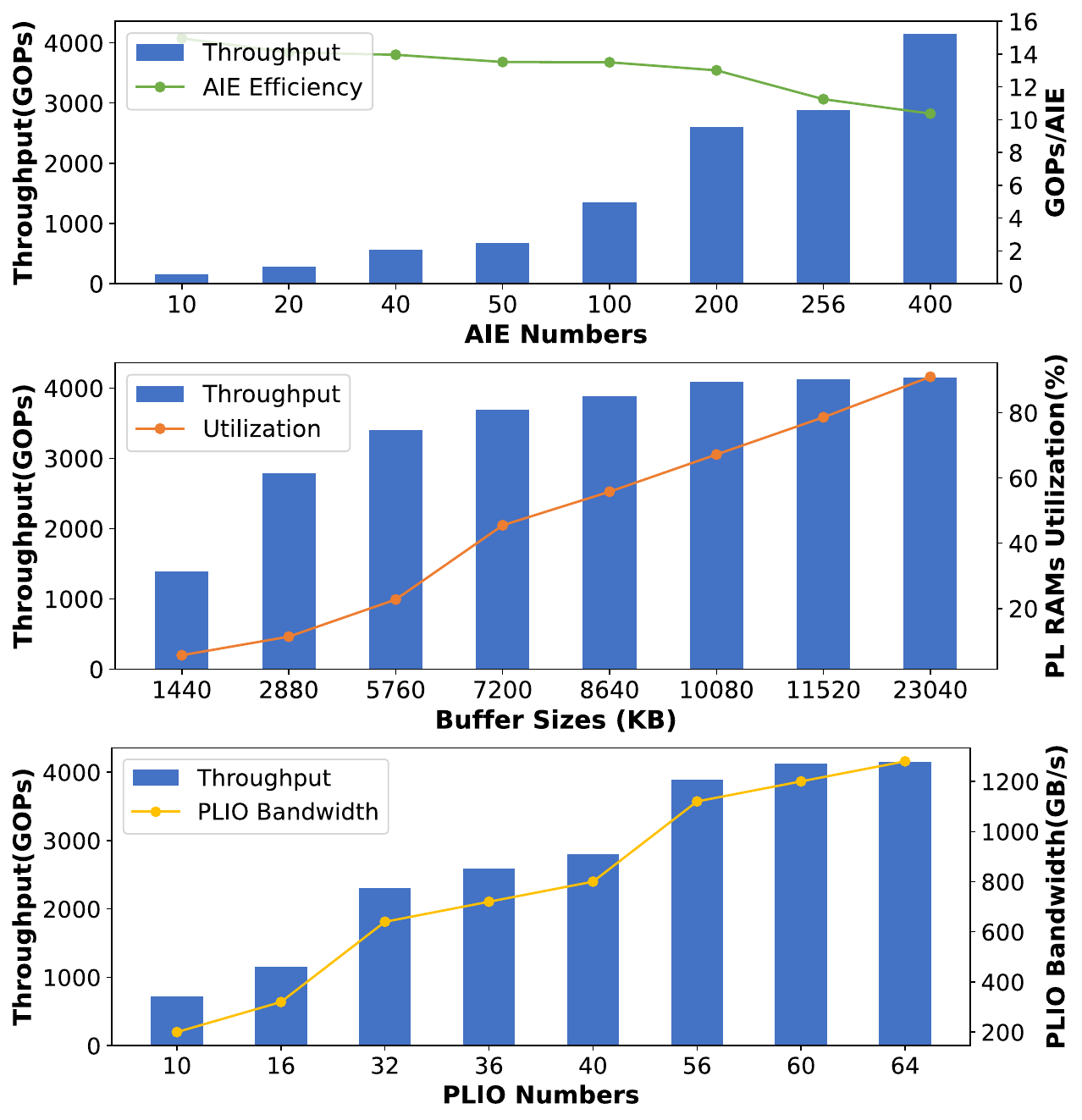}
    \caption{Throughput Evaluation of Different AIE Numbers, PLIO Numbers, and PL Buffer Sizes}
    \label{fig:factors}
    \vspace{-0.6cm}
\end{figure}

We evaluate the scalability of WideSA while increasing AIE utilization and analyze how various factors influence performance. 
The results, presented in Figure~\ref{fig:factors}, show a significant increase in throughput as the number of AIEs increases.
In addition, the AIE efficiency results demonstrate that our approach scales effectively from small-scale to large-scale designs. 
However, when the number exceeds 200, the efficiency of a single AIE core decreases due to the memory-bound condition caused by the number of PLIOs and the size of the PL buffer.
The increase in PLIO numbers and buffer sizes leads to increased throughput, suggesting that enhancing the bandwidth between different fabrics of ACAP can improve performance. 
This indicates that managing the resources and data flow between different components of the ACAP is crucial for achieving better performance.

\section{Conclusion}
In this paper, we present a high array utilization mapping scheme for uniform recurrences on the Versal ACAP architecture.
Additionally, we propose several optimizations aimed at enhancing overall performance within an automatic mapping framework. 
Through extensive evaluations using typical benchmarks and diverse data types, we assess the efficiency of the WideSA framework. 
In the future work, we aim to integrate WideSA into the MLIR-AIE workflow and develop an end-to-end compilation tool that incorporates automatic design space exploration.

\section*{Acknowledgement}

This work was partly supported by the National Natural Science Foundation of China (Grant No. 62090021) and the National Key R\&D Program of China (Grant No. 2022YFB4500500).

\bibliographystyle{IEEEtran}
\bibliography{ref.bib}

\end{CJK*}
\end{document}